\documentclass[api,reprint,amsmath,amssymb,superscriptaddress,showpacs,floatfix]{revtex4-1}
\usepackage{graphicx}
\usepackage{dcolumn}
\usepackage{bm}
\usepackage[colorlinks, allcolors=blue]{hyperref}
\usepackage[usenames, dvipsnames]{color}
\usepackage{setspace}
\usepackage{scalerel}
\usepackage{stackengine}
\usepackage{comment}
\usepackage{amsmath}
\stackMath
\newsavebox\tmpbox

\begin{document}
\title{Quantum phases of spin-1 system on 3/4 and 3/5 skewed ladders}

\author{Sambunath Das}
\email{sambunath.das46@gmail.com}
\affiliation{Solid State and Structural Chemistry Unit, Indian Institute of Science, Bangalore 560012, India}

\author{Dayasindhu Dey}
\email{dayasindhu.dey@gmail.com}
\affiliation{Solid State and Structural Chemistry Unit, Indian Institute of Science, Bangalore 560012, India}

\author{S. Ramasesha}
\email{ramasesh@iisc.ac.in}
\affiliation{Solid State and Structural Chemistry Unit, Indian Institute of Science, Bangalore 560012, India}

\author{Manoranjan Kumar}
\email{manoranjan.kumar@bose.res.in}
\affiliation{S. N. Bose National Centre for Basic Sciences, Block - JD, Sector - III, Salt Lake, Kolkata 700106, India}

\date{\today}

\begin{abstract}
        We study the quantum phase transitions of frustrated antiferromagnetic
	Heisenberg spin-1 systems on the 3/4 and 3/5 skewed two leg ladder
	geometries. These systems can be viewed as arising by periodically
	removing rung bonds from a zigzag ladder. We find that in large systems,
	the ground state (gs) of the 3/4 ladder switches from a singlet to a
	magnetic state for $J_1 \ge 1.82$; the gs spin corresponds to
	ferromagnetic alignment of effective $S = 2$ objects on each unit cell.
	The gs of antiferromagnetic exchange Heisenberg spin-1 system on a 3/5 skewed
	ladder is highly frustrated and has spiral spin arrangements. The
	amplitude of the spin density wave in the 3/5 ladder is significantly
	larger compared to that in the magnetic state of the 3/4 ladder. The gs
	of the system switches between singlet state and low spin
	magnetic states multiple times on tuning $J_1$ in a finite size system.
	The switching pattern is nonmonotonic as a function of $J_1$,
	and depends on the system size. It
	appears to be the consequence of higher $J_1$ favoring higher spin magnetic
	state and the finite system favoring a standing spin wave. For some
	specific parameter values,  the magnetic gs in the 3/5 system is doubly
	degenerate in two different mirror symmetry
	subspaces. This degeneracy leads to spontaneous spin parity and mirror
	symmetry breaking giving rise to spin current in the gs of
	the system.
\end{abstract}

\maketitle

\section{\label{sec:intro}Introduction}
Spin chains and ladders show strong quantum fluctuations due to spatial 
confinement and are extensively studied to explore the intriguing magnetic 
properties of these systems. The excitations and ground state (gs) properties
of these systems depend on the magnitude of the spins at the lattice sites 
(half-odd-integer or integer)~\cite{haldane83a,*haldane83b,aklt87,aklt88} 
and topology of the exchange interactions~\cite{ckm69a,*ckm69b,white96,mk2015,
soos-jpcm-2016,hamada88,chubukov91,chitra95,itoi2001,meisner2006,meisner2007,
kecke2007,vekua2007,mahdavifar2008,dmitriev2008,hikihara2008,sudan2009,
meisner2009,sirker2010,mk_bow,mk2012,aslam_magnon}. 
A Heisenberg antiferromagnetic(HAF) spin-1/2 chain with only nearest neighbor 
exchange interaction, $J_1$, shows a quasi-long range order in the gs and a 
gapless spectrum~\cite{mikeska2004}, whereas a spin-1/2 HAF chain with nearest neighbor and next 
nearest neighbor exchange interactions $J_1$ and $J_2$, respectively, can show 
quasi-long range or short range order and gapless or gapped spectrum depending on 
the ratio $J_2/J_1$~\cite{ckm69a,*ckm69b,white96,mk2015,soos-jpcm-2016}. 
The HAF spin-1 chains, on the other hand, with only 
nearest neighbor-exchange interaction have short range order in the gs and
gapped spectrum as pointed out by Haldane~\cite{haldane83a,*haldane83b}, and
gs can be represented as a valance bond solid
(VBS)~\cite{aklt87,aklt88,schollwock96} which is of the same universality class
as the Affleck, Kennedy, Lieb and  Tasaki (AKLT) states~\cite{aklt87,aklt88}.
The AKLT state has inspired many numerical techniques like matrix product
states~\cite{rommer95,verstraete2008,schollwock2011}, Tensor Network
methods~\cite{orus2014} and projected entangled pair states (PEPS)
methods~\cite{murg2007,verstraete2008}. AKLT states can also be
represented as cluster states which can be used in measurement-based quantum
computation~\cite{cirac2004,affleck2011}.

\begin{figure}
\includegraphics[width=\columnwidth]{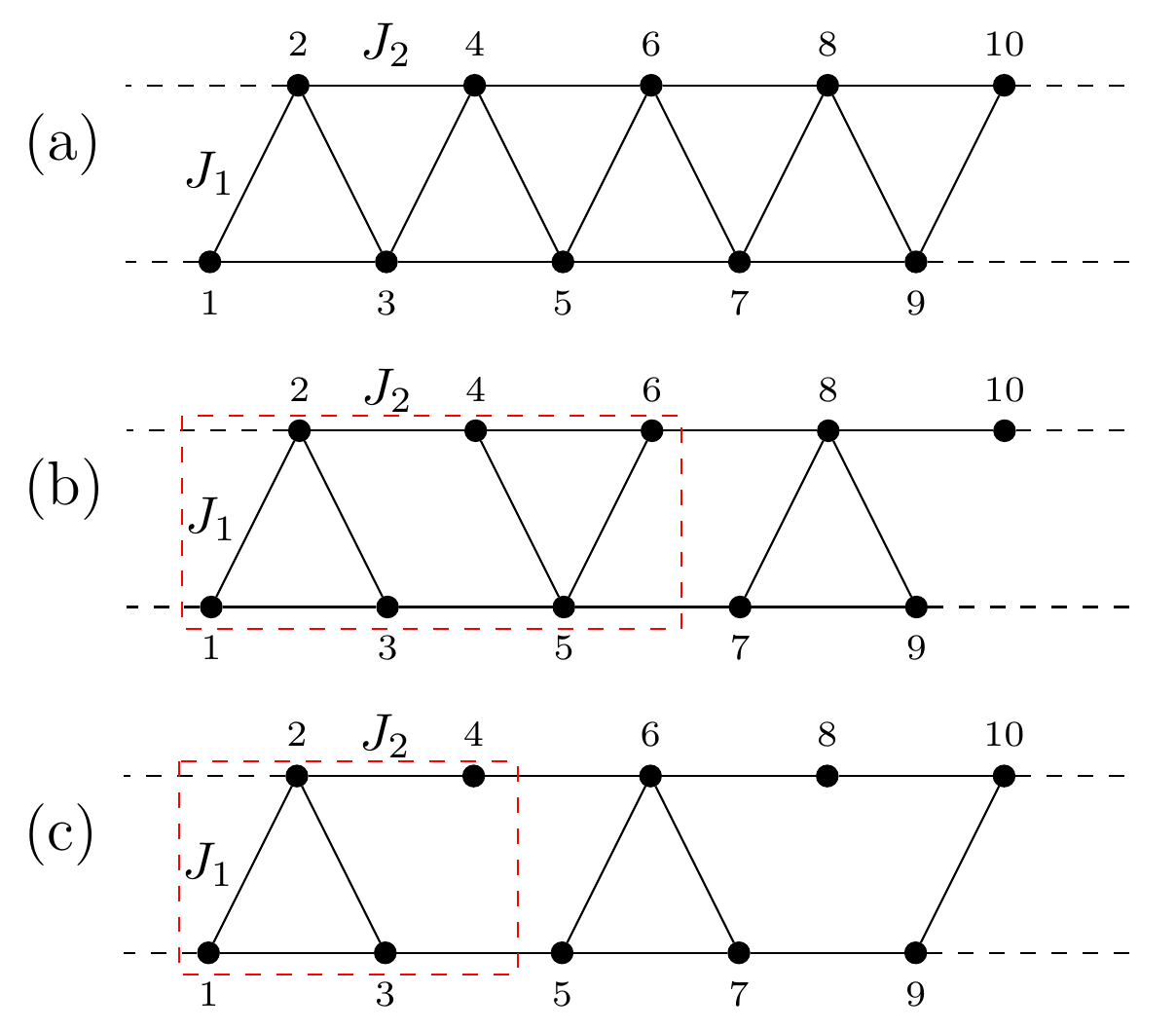}
\caption{\label{fig1} Schematic diagram of (a) zigzag ladder, (b) 3/4 skewed
	ladder and (c) 3/5 skewed ladder. The unit cell of each structure is
	shown inside the box in red.}
\end{figure}

The first experimental realization of the spin-1 Haldane system was in the well
known transition metal chain compound Ni(C$_2$H$_8$N$_2$)$_2$NO$_2$ClO$_4$
(NENP)~\cite{renard87,renard88,katsumata89}. The HAF spin-1 chain exhibits
topological phase with, spin-1/2 edge modes leading to four fold degenerate gs
in  thermodynamic limit. In spin-1 chain, the gs correlation length, $\xi =
6.05$ lattice constant and a large  spin gap to the excited state,
$\Delta_{\rm{ST}} \sim 0.41 J_1$~\cite{dayaprb16,white-huse-prb93}. The gs of a
two leg HAF spin-1 normal ladder  is a plaquette-singlet solid state (PSSS)
where two spin-1/2 singlet dimer are sitting at each rung and there is no
overlap to the VBS state in the large rung exchange limit~\cite{todo2001}. On a
zigzag ladder, the gs of a spin-1 HAF model is the Haldane phase in the weak rung
interaction limit, while a double Haldane state is the gs in the large rung
interaction limit~\cite{hikihara2002,natalia2016}. Another class of spin ladders
that have been studied in recent times is the skewed ladder which contains
slanted rung bonds which periodically displace rung bonds as well as
periodically missing rung bonds of the original ladder (Fig.~\ref{fig1}). This
leads to fused equivalent or inequivalent cyclic rings. An example of such a skewed
ladder is the 5/7 skewed ladder with alternately fused five- and seven-membered
rings while 5/5 skewed system consists of equivalent five membered rings fused
together. The 5/7 system has been studied in both the spin-1/2~\cite{thomas2012,geet, 57plateau} and spin-1
cases~\cite{sambu57}. Similarly, the 3/4, 3/5 and 5/5 spin-1/2 skewed 
ladders have also been studied earlier and it was shown that 
the gs of hetero skewed ladders (polygons of unequal number of vertices fused 
together), such as 5/7 and 3/4 are magnetic~\cite{geet}. In the large rung exchange 
limit the gs wavefunction of 3/4 skewed ladder can be represented as a product 
of rung singlet dimers and one ferromagnetically interacting spin-1/2 object per
triangle~\cite{geet}. The gs of a spin-1/2 system on a 3/5 geometry is a low 
spin magnetic state for intermediate values of $J_1$, but evolves to a 
gapless antiferromagnetic state for large $J_1$~\cite{geet}. As outlined above, the 
spin-1/2 and spin-1 systems exhibit fundamentally different behavior in case of 
the HAF spin chains. This has prompted us to study the spin-1 system on these ladders, to 
explore the quantum phase transitions in these systems as the ratio of the rung 
exchange to the ladder exchange strengths is varied.

In this paper, we focus on the gs properties 
of spin-1 objects arranged on 3/4 and 3/5 skewed ladders interacting via 
competing rung and leg antiferromagnetic exchange interactions of strength $J_1$ 
and $J_2$ respectively. The structure of zigzag, 3/4 and 3/5  skewed ladders are 
shown in Fig.~\ref{fig1}(a)--\ref{fig1}(c). In section~\ref{sec2} we discuss the 
model Hamiltonian and numerical methods used in the paper. The results for 3/4 and 3/5 skewed 
ladders are discussed in section~\ref{sec3}. In section~\ref{sec3a}, we show that 
the gs of the 3/4 system switches from a singlet to a magnetic state for 
$J_1 \geq 1.82$ with magnetization per unit cell `$\langle m \rangle$' taking a value of $2$. 
In section~\ref{sec3b}, we show that the gs of 3/5 ladder exhibits short 
range spin correlations and a spiral spin density wave, the spin of the gs depends 
both on size of the system and $J_1$ for intermediate $J_1$ values; the gs is 
magnetic with one unpaired spin per unit cell in the large $J_1$ limit. For some
specific $J_1$ values, the gs is degenerate and exhibits vector chiral phase due 
to simultaneous breaking of spin inversion and reflection  symmetries. Section~\ref{sec4}
provides summary of results and conclusions.

\section{\label{sec2}Model and method}
We consider the HAF spin-1 model on 3/4 and 3/5 skewed ladders shown in 
Fig.~\ref{fig1}(b) and (c) respectively; all  exchange  interactions  between  
the  neighboring  spins  are antiferromagnetic in nature. The site numberings 
are shown in Figs.~\ref{fig1}(b) and~\ref{fig1}(c). The exchange interaction 
along rung and leg bonds are represented as $J_1$ and $J_2$ respectively. The 
exchange interaction $J_2$ is set to 1, to define the energy scale in all the 
studies. The model Hamiltonian of 3/4 skewed ladder can be written as
\begin{eqnarray}
	H_{3/4} &=& J_1 \sum_{k=1}^{N/3} \vec{S}_{3k-1} \cdot \left(\vec{S}_{3k-2} 
	+ \vec{S}_{3k}\right) \nonumber \\
	& & \qquad + J_2 \sum_{k=1}^{N-2} \vec{S}_{k} \cdot\vec{S}_{k+2}
\label{eq:ham34}
\end{eqnarray}
where the system consists of $N=6n$ sites, with $n$ being the number of unit 
cells and an open boundary condition is assumed (Fig.~\ref{fig1}(b)). The first 
term denotes the rung exchange, the second denotes the exchange interactions 
along the legs. The model Hamiltonian of 3/5 skewed ladder can be written as
\begin{eqnarray}
	H_{3/5} &=& J_1 \sum_{i=1}^{n} \vec{S}_{4i-2} \cdot 
	\left(\vec{S}_{4i-3} + \vec{S}_{4i-1}\right) \nonumber \\
	& & \qquad + J_1 \, \vec{S}_{4n+2} \cdot \vec{S}_{4n}
	+ J_2 \sum_{i=1}^{4n} \vec{S}_{i} \cdot\vec{S}_{i+2}.
\label{eq:ham35}
\end{eqnarray}
Where $n$ is the number of unit cells and an open boundary condition is assumed. 
In case of periodic boundary condition, the Hamiltonian is modified accordingly and 
sites $4i-2$ and $4i$ are the inversion centres of the system.  Both $H_{3/4}$ 
and $H_{3/5}$ conserve total spin $S^2$, and $z$-component of the total spin
$S^z$.

We use exact diagonalization (ED) technique for solving these ladders with up to 
16 spins with periodic boundary condition (PBC). For larger system sizes, we use 
the now well known finite density matrix renormalization group (DMRG)
method~\cite{white-prl92,white-prb93,schollwock2005,karen2006} which retains the
size of the Hamiltonian matrix at all system sizes. This is achieved by a
systematic truncation of the irrelevant degrees of freedom by selecting `$m$'
states with largest eigenvalues of the density matrix to span the Fock space of a
part of the full system. The chosen value of $m$ is up to $500$ in our studies
and it keeps the truncation error of the density below $\sim 10^{-10}$. We also 
carry out 6-10 finite sweeps for convergence. For some values of $J_1$ and for 
large system sizes we have employed $m = 1600$ and 12 finite DMRG sweeps to 
improve convergence. The largest system size used in this 
paper is $N=98$ or 24 unit cells with open boundary condition.

\section{\label{sec3}Results and discussions}
In this section we present the results for both 3/4 and 3/5 skewed ladders. The gs 
of the 3/4 skewed ladder shows a transition from a singlet to a magnetic phase,  
whereas the gs of 3/5 ladder has spiral spin arrangement and switches between
different magnetic and singlet states upon tuning $J_1$. In the decoupled limit the gs of 
both systems show Haldane phase, whereas, in large $J_1$ limit the gs exhibits 
strong rung trimer formation. In strong rung coupling limit (very large $J_1$) the 
gs of the system has an effective spin-1 on each triangle with those triangles 
interacting ferromagnetically. To analyze the magnetic transitions, various 
quantities like gs energies $E_{gs}$, energy gaps to low-lying excited states 
$\Gamma_l$, bond orders $b_{i,j}$ between sites $i$ and $j$ and local spin densities 
$\rho_i$ at site $i$ are analyzed, as a function of $J_1$.  The excitation 
gaps $\Gamma_l$ are defined as 
\begin{equation}
     \Gamma_l = E_{0} (S^{z}=l) - E_{0} (S^{z}=0)
\label{eq:gamma}
\end{equation}
where $E_{0} (S^{z}= l)$  and $E_{0} (S^{z}= 0)$  are  the energies of the
lowest states in the $S^{z} = l$ and $S^{z}=0$ manifold, respectively, and $l$ 
is an integer.
The model Hamiltonians in Eqs. \ref{eq:ham34} and \ref{eq:ham35} are 
isotropic and therefore, conserve  total spin $S$ and its $z$-component $S^z$.
For a magnetic gs with spin $S$, $E_0(S^{z}=S)$ is degenerate
with $E_0(S^z=m)$ where $-S \leq m \leq S$ and the state is $(2S+1)$ fold
degenerate. Thus $S$ is the gs spin if the lowest energy level in every 
$S^{z} \leq S$ are the same or $\Gamma_l =0$ for all $l \leq S$. The bond order  
$b_{ij} = -\langle \vec{S}_i \cdot \vec{S}_{j} \rangle$, spin density 
$\rho_i = \langle {S}_i^{z} \rangle$ and correlation function  
$C(r)= \langle \vec{S}_i \cdot \vec{S}_{i+r} \rangle$  are the gs
expectation values calculated to characterize the gs with $S^{z}= S$.  Due to
symmetry there are only two unique sites and three unique bonds per unit cell in
the 3/4 ladder, whereas there are three unique sites and four bonds in the 3/5 
ladder. Therefore, it is sufficient to obtain all the above quantities for these
unique sites and bonds.

\subsection{\label{sec3a}3/4 Skewed ladder}
\begin{figure}
\includegraphics[width=\columnwidth]{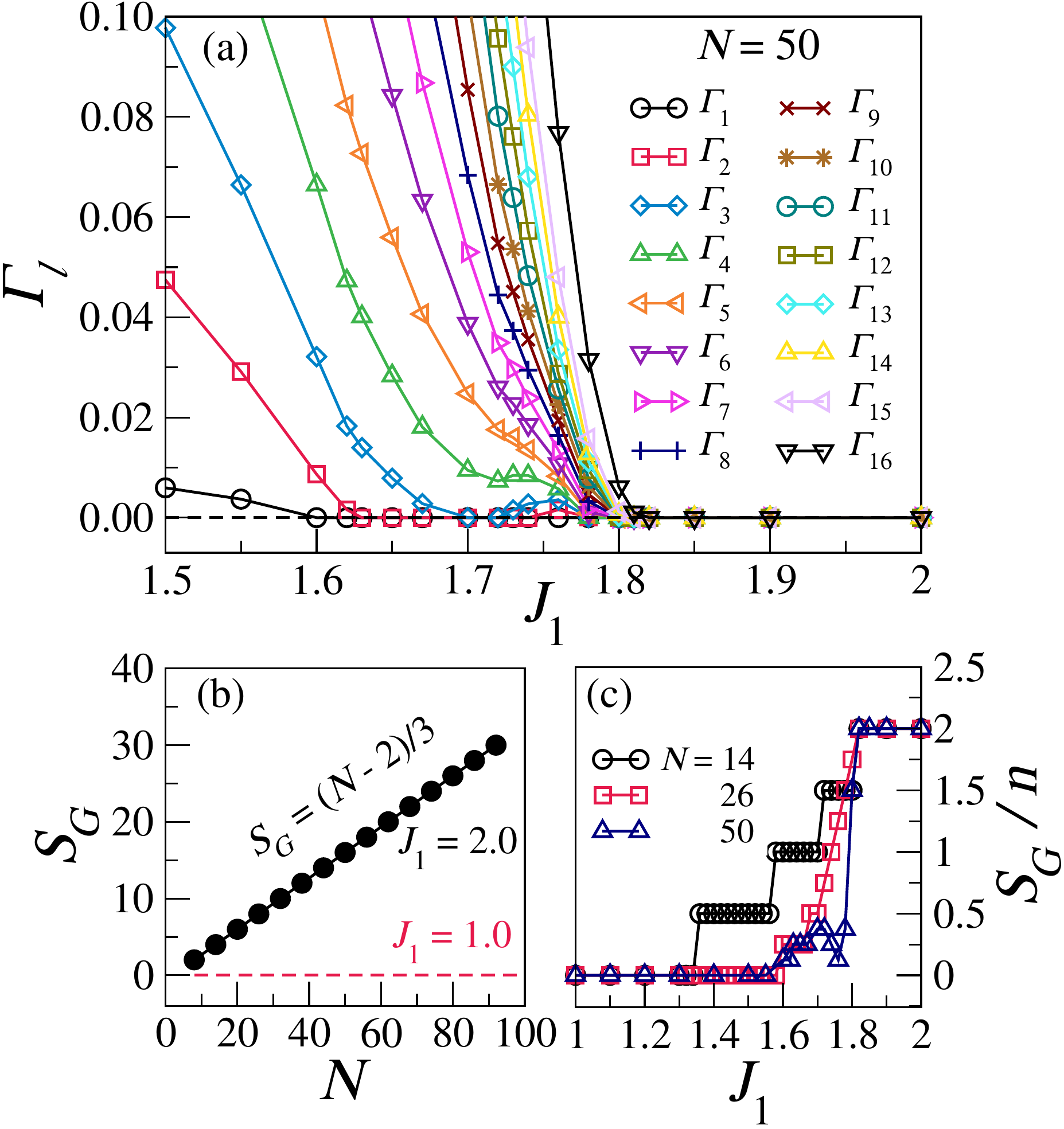}
\caption{\label{fig2}(a) The lowest excitation gaps $\Gamma_l$ are shown as 
	function of $J_1$ for 3/4 ladder of 50 sites with open boundary 
	condition. For $J_1 < 1.6$, all the $\Gamma_l$ are non-zero, whereas, 
	for $1.6 < J_1 < 1.82$, the gs transitions to a magnetic state. For 
	$J_1 \geq 1.82$, the gs spin equals twice the number of unit cells in the 
	system. (b) The $S_G$ is shown as a function of system sizes for 
	$J_1=1.0$ and $2.0$. Each data point corresponds to integer number of 
	unit cells. (c) $S_G$ vs. $J_1$ for different system sizes.}
\end{figure}

The magnetic gs of the 3/4 skewed ladder is obtained using excitation gaps 
$\Gamma_l$, defined in Eq.~(\ref{eq:gamma}). In this system the gs rapidly evolves 
from a singlet state to a magnetic state with spin $S=2n$ where $n$ is the 
number of unit cells in the system. This transition seems continuous in systems 
with open boundary condition (OBC) with system size $N=50$. In this system
(Fig.~\ref{fig2}(a)) the gs spin, $S_G$, changes gradually beyond $J_1
= 1.67$ and near $J_1 \sim 1.75$, $S_G$ increases rapidly and beyond $J_1 =
1.82$, $S_G$ achieves the highest gs spin of $S_G = 16$.
The transition region  shrinks dramatically to  
$1.65 < J_1 < 1.67$ for $N=18$ with periodic boundary condition (PBC) and for 
both PBC and OBC case the transition region decreases with system size.
The scaling of the total spin $S_G$ of the gs is shown as function of unit cell 
in two parameter regimes in Fig.~\ref{fig2}(b). The gs remains a singlet for 
$J_1 < 1.6$, whereas, for larger $J_1 \geq 1.82$, $S_G$ vs $N$ shows a linear variation 
with slope 1/3. The linear variation for $S_G$ vs $N$ for $J_1=2$ is shown in 
Fig.~\ref{fig2}(b). The finite size effect of $S_G$ is shown in $S_G$ vs $J_1$ curve 
for three system sizes $N=14$, 26 and $50$ in Fig.~\ref{fig2}(c). We 
notice that the $S_G$ increases in a stepwise fashion for small system size but 
shows a sharp  increase as the system size increases.

\begin{figure}
\includegraphics[width=\columnwidth]{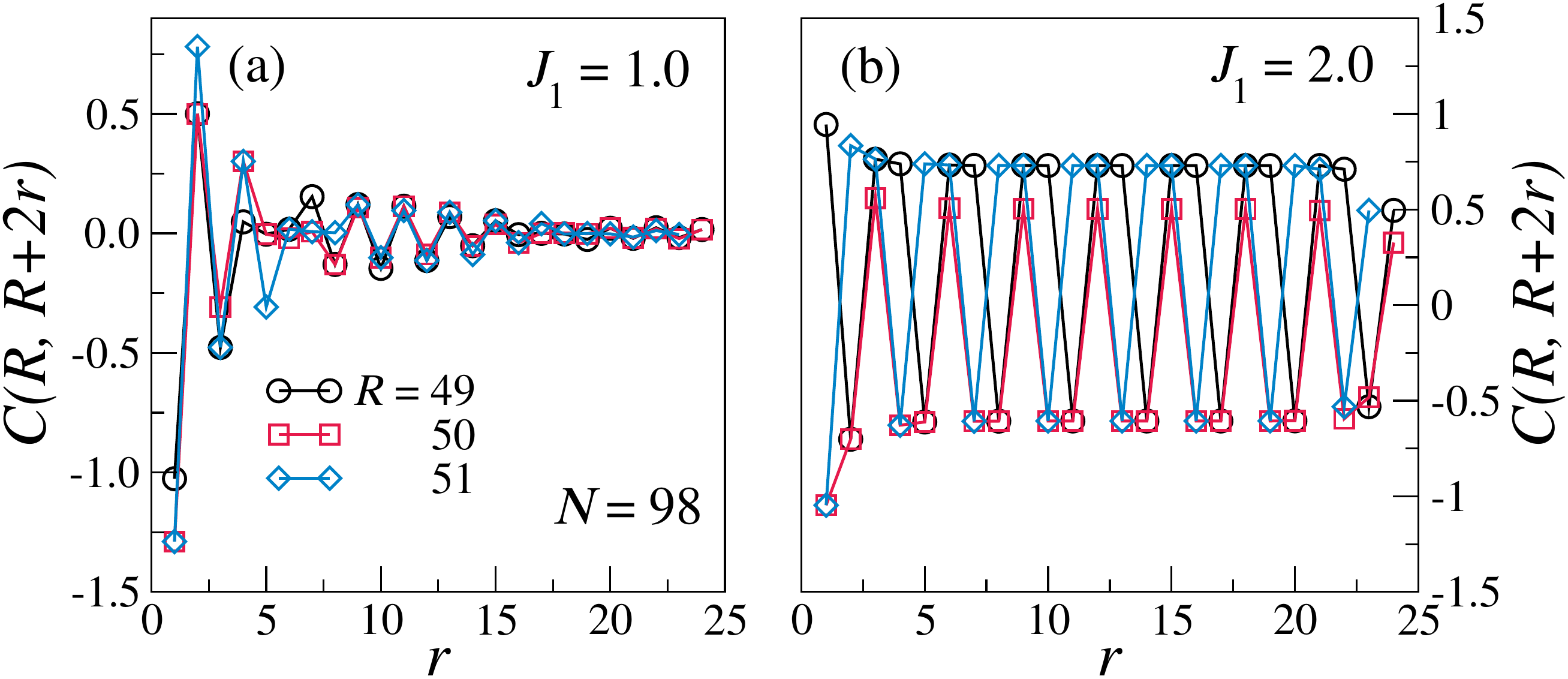}
\caption{\label{fig3} The spin-spin correlations between the spins in the lower leg (R = 49 and R=51),
spins in upper leg (R = 50) for a 3/4 ladder with N=98 spins with OBC with (a) $J_1=1.0$, the singlet
	regime and  (b) $J_1 = 2.0$, the magnetic regime.}
\end{figure}

\begin{figure}
\includegraphics[width=\columnwidth]{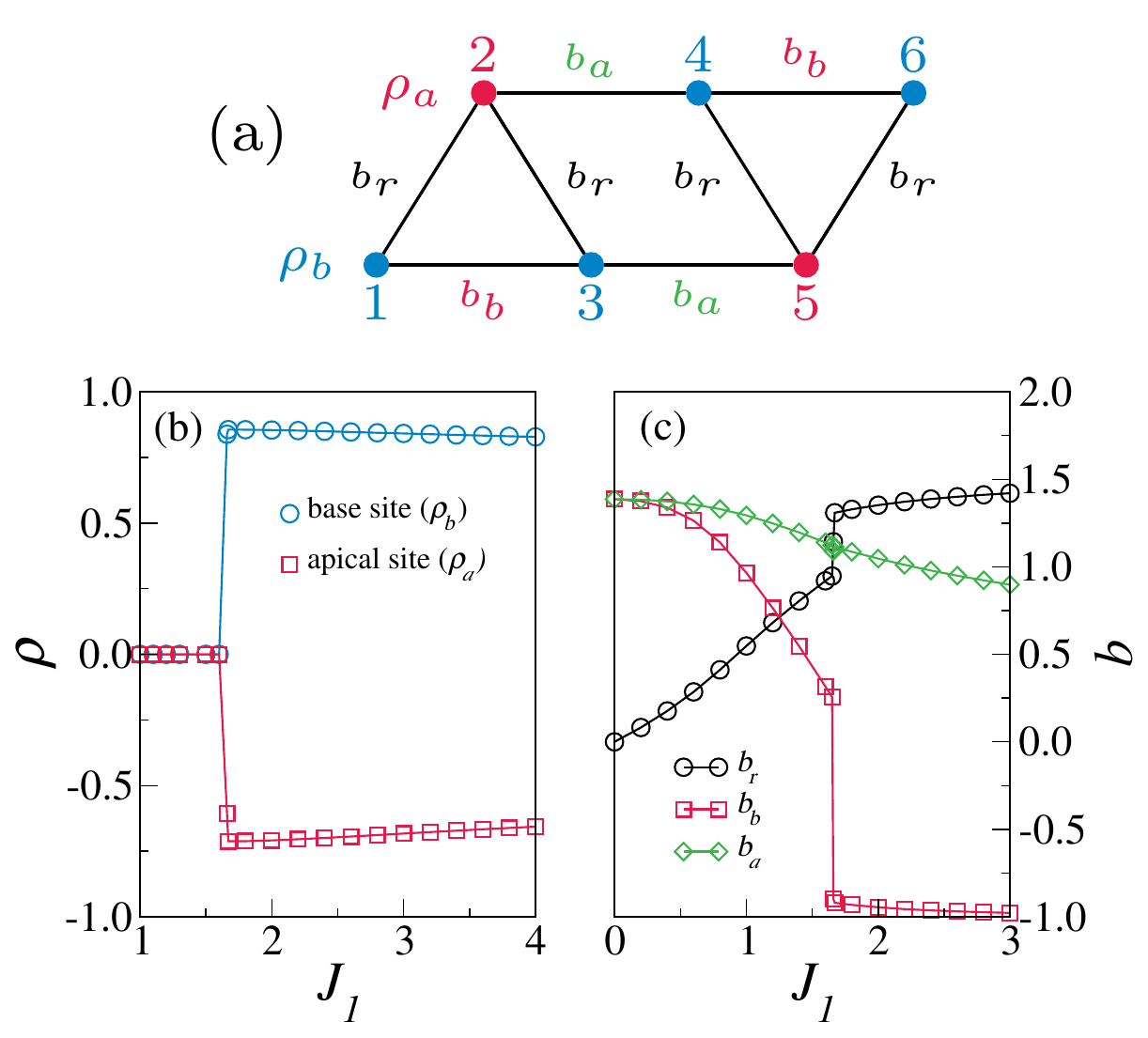}
	\caption{\label{fig4}(a)The base and the apical spins are shown in blue and red;
	three different types of bonds namely, the rung bonds ($b_r$), the basal
	bonds ($b_b$) and the apical bonds ($b_a$) are shown for a unit cell 
	(b) Spin densities $\rho_b$ and $\rho_a$ for the base sites and the 
	apical sites, respectively, as a function of $J_1$ (c) bond orders 
	$b_r$, $b_b$ and $b_a$ as a function of $J_1$ are shown.}
\end{figure}

In small $J_1$ limit, spin correlation $C(r)$ behavior is similar to that of a 
HAF spin-1 chain with short correlation length.  In Fig.~\ref{fig3}(a) and~\ref{fig3}(b) 
total correlation function $C(R, R+2r)$ is shown for $J_1=1.0$ and 2.0 respectively. 
There are two unique sites one at the base of the triangle and other at the apex of 
the triangle. We consider reference sites on the middle triangle of the ladder 
with site labels 49, 50 and 51 of a $N=98$ site system. Site 49 and 51 are at 
the base of the triangle whereas site 50 is at the apex of the triangle, and 
correlations from these points $C(49, 49+2r)$ and $C(51, 51+2r)$ are the 
correlations between the reference sites on the lower leg of the ladder with all 
the sites on the lower leg, whereas $C(50, 50+2r)$ are the correlations between 
the reference site on the upper leg of the ladder with all the other sites on 
the upper leg. The $C(R,R+2r)$ for $R=49$ and 51 sites are similar as both 
these sites belong to the base of the triangle and are equivalent by symmetry. In 
Fig.~\ref{fig3}(a) $C(R, R+2r)$ for $J_1=1.0$ seems to show oscillatory behavior 
with exponentially decreasing amplitudes. This seems to indicate existence of 
spin wave packets in the singlet state. The $C(R, R+2r)$ in magnetic regime 
$J_1=2.0$ are shown Fig~\ref{fig3}(b), and spins are antiferromagnetically aligned 
and have long range correlations. In this regime, in the bulk of the system, the 
correlations are oscillating with almost constant amplitude. The correlation 
between the reference spins in the base of the triangle is always ferromagnetic 
with spins at the base of the triangle and antiferromagnetic with the spin at 
the apex of the triangle. This is true on both legs. We have also computed the 
spin densities in the gs of the ladder as a function of $J_1$. The 
spin densities vanish at all the sites in the singlet gs. We note from 
Fig.~\ref{fig4}, that the apex and the base sites have opposite spin densities 
in the magnetic gs. The spin densities at all the sites are nearly 
constant. Indeed, the spin-spin correlations in the magnetic state are well 
approximated by the product of the spin densities at the corresponding sites.

In large $J_1$ limit the spin densities show that the gs behavior can be 
thought of as each triangle having effective spin-1, which are
interacting ferromagnetically. The spin density of an isolated triangle with large 
$J_1$ interaction is also $0.75$ on the basal sites and $-0.5$ at the apex which 
is close to the values found in the gs at large $J_1$.

In 3/4 system there are only three distinct bond orders corresponding to the rung 
bond $b_r$, bond between the basal sites $b_b$ and bond between the apical and 
base sites $b_a$. We notice that both types of bonds on the legs ($b_a$ and
$b_b$) show decreasing 
bond order as $J_1$ increases and the basal bond order shows jump at $J_1=1.65$. 
Whereas, the magnitude of rung bond order increases continuously with $J_1$ and 
jumps suddenly at $J_1=1.65$. The jump in the rung bond order is due to transition 
of the gs from a singlet to a magnetic state. The saturation value of the three 
bonds for large $J_1$ limit is $+1.5$, $+1.0$ and $-1.0$ respectively. The value 
$b_r=+1.5$ corresponds to two spin-1/2 objects separately forming bonds and total bond
order is twice of a singlet bond order $+3/4$. The $b_b=+1$ and $b_a=-1$ 
corresponding to the ferromagnetic bond formation between axial and basal sites
and intermediate spin state between the basal sites. In small 
$J_1$ limit, the $b_r$ rung bond is weak and $b_r$ is close to zero. Whereas,  
$b_b$ and $b_a$, for small values of $J_1$,  start with value close to the bond order 
of a spin-1 chain $\sim 1.40$, they  decrease with increasing $J_1$; $b_b$ becomes ferromagnetic 
and $b_a$ remains antiferromagnetic.

To understand the spin density and bond 
order behavior in large $J_1$ limit, we analyze a system with three spins on a 
triangle, and exchange interaction of the two sides and base are  $J_1$ and  
$J_2=1$, respectively. The gs is in spin-1 states and  there are six possible 
spin configurations. For $J_1 \geq 2$  we calculate the gs wavefunction 
of the system in the $S=1$ and $S^z = 1$ manifold as
\begin{eqnarray}
	\mid \Psi &(& S=1, S^z = 1) \rangle \nonumber \\
	& = & \sqrt{\frac{3}{5}} \left[\mid 1,-1,1 \rangle 
	+ \frac{1}{2} (\mid 0,0,1 \rangle + \mid 1,0,0 \rangle) \right. \nonumber \\
	& & \left. -\frac{1}{6} (\mid -1,1,1 \rangle + \mid 1,1,-1 \rangle)
	-\frac{1}{3} \mid 0,1,0 \rangle \right].
\end{eqnarray}
We notice that the large contribution ($60\%$) is from the state $\mid 1,-1,1 \rangle$ 
and a smaller ($15\%$) contribution is from  the linear combination $1/2(\mid{0,0,1}\rangle+\mid 1,0,0 \rangle)$ 
spin configuration.  The sites 1 and 3 are symmetric and total contribution of spin 
density comes from these two configurations, and $60\%$ contribution to $\langle
S^z \rangle$ arises from $\mid 1,-1,1 \rangle$ while a contribution of $15\%$ to
$\langle S^z \rangle$ arises from $(\mid 0,0,1 \rangle + \mid 1,0,0 \rangle)$,
resulting in a total spin density of $0.75$. The configuration $\mid 1,-1,1 \rangle$ contributes $-0.6$ 
to $\langle S^z \rangle$ at site $2$ while the states 
$(\mid -1,1,1 \rangle + \mid 1,1,-1 \rangle)$ and $\mid 1,-1,1 \rangle$ contribute 
$+0.1$ to $\langle S^z \rangle$ at site $2$. This results in total spin density
at site 1 and 2 of $0.75$ and $-0.5$ respectively in large $J_1$ ($J_1 \geq
2.0$) limit. The contribution to  the 
$z$ component of all three bond orders mostly arises from the configuration 
$\mid 1,-1,1 \rangle$ and the bonds $1-2$ and $2-3$ are singlet in nature whereas 
the bond $1-3$ is ferromagnetic.

\subsection{\label{sec3b}3/5 Skewed ladder}
The 3/5 ladder has 4 spins per unit cell as shown schematically in 
Fig.~\ref{fig1}(c) and in each unit cell three sites form a triangle, whereas 
$4^{\rm{th}}$ site is connected to the apex of the triangle. There is one rung 
bond at each odd-numbered site and two rung bonds at alternate even numbered 
sites, $(4k - 2)$, $k = 1$, 2, $\cdots$. With periodic boundary condition, the 
sites $2k$ at the apices of triangles and pentagons are inversion centres.
The model Hamiltonian of this system is shown in Eq.~(\ref{eq:ham35}) 

\begin{figure}
\includegraphics[width=\columnwidth]{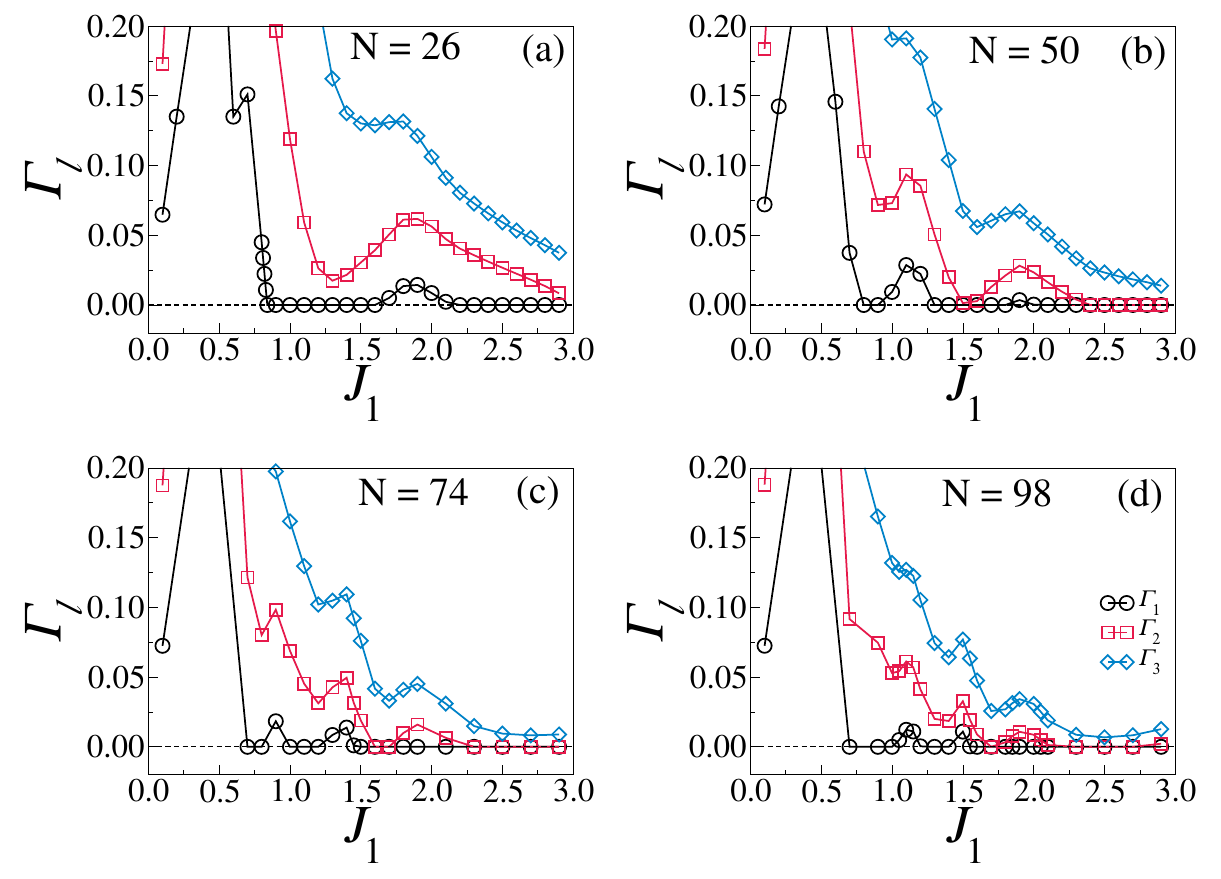}
\caption{\label{fig5}The lowest excitation energy gaps in different $S^z$
	sectors, $\Gamma_l$ for different $l=S^z$ manifolds as a function of the
	rung exchange interaction $J_1$ for systems of (a) $N = 26$ spins (6 unit
	cells), (b) $N = 50$ spins (12 unit cells), (c) $N = 74$ spins (18 unit
	cells), and  (c) $N = 98$ (24 unit cells) with open boundary
	condition. For larger system sizes, the gs switches between
	magnetic and singlet states as $J_1$ is increased.}
\end{figure}

The gs of the 3/5 system is singlet in low $J_1$ limit and toggles between 
singlet and different magnetic states as $J_1$ is increased.
We find that the evolution of the spin of the gs depends not only on
$J_1$ but also on the system size. We show this dependence for four
representative sizes of the systems with open boundary condition. For the
system with 6 unit cells (Fig.~\ref{fig5}) ($N=26$), the gs
switches from a singlet to a triplet at $J_1 = 0.84$ and it then switches back
to a singlet at $J_1 = 1.7$. The triplet then becomes a gs from $J_1 =
2.15$ to $J_1 = 3.0$. Beyond $J_1 = 3.0$ the gs is a quintet until
$J_1 = 6.5$, a septet ($S = 3$) from $J_1 = 6.5$ to $7.0$ a nonet ($S = 4$) from
$J_1 = 7.0$ to $8.5$ and $S = 5$ beyond $J_1 = 8.5$. The highest spin in the $N
= 26$ systems with 6 unit cells can be $S = 6$, corresponding to ferromagnetic
arrangement of spins at sites $4k$. In large systems these transitions occur at
different values of $J_1$ (Fig.~\ref{fig5}).

\begin{figure}
\includegraphics[width=\columnwidth]{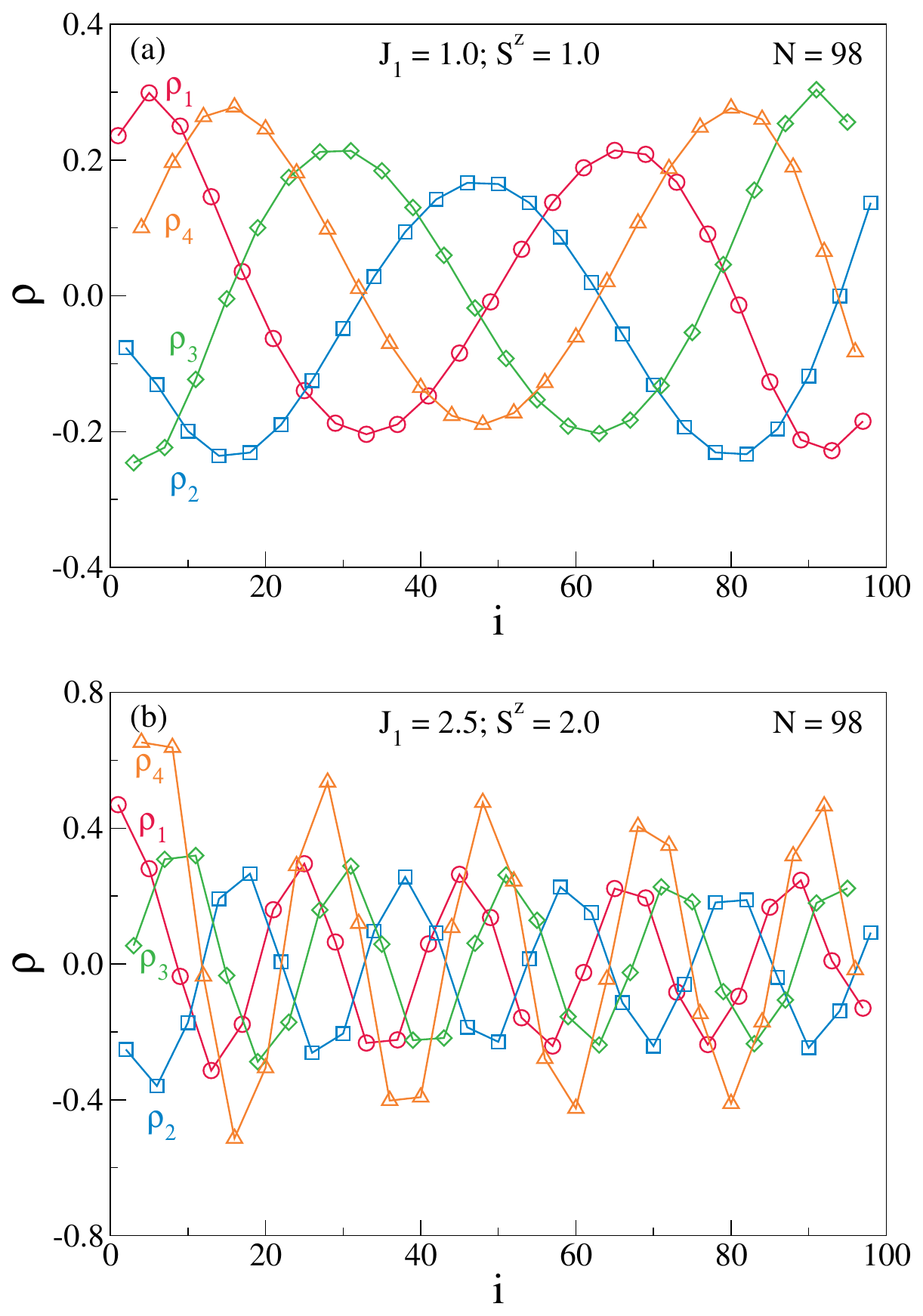}
\caption{\label{fig6}Spin densities of a 3/5 ladder of $N = 98$ spins (OBC).
	(a) spin densities for $J_1=1.0$ in the triplet gs are shown
	with continuous red, blue, green and orange curve (the eye guide) for 1,
	2, 3 and 4 sites in every unit cells. (b) spin densities for $J_1=2.5$
	are shown in the quintet gs. We have chosen $S^z = S_G$ in
	both the cases.}
\end{figure}

\begin{figure}
	\includegraphics[width=\columnwidth]{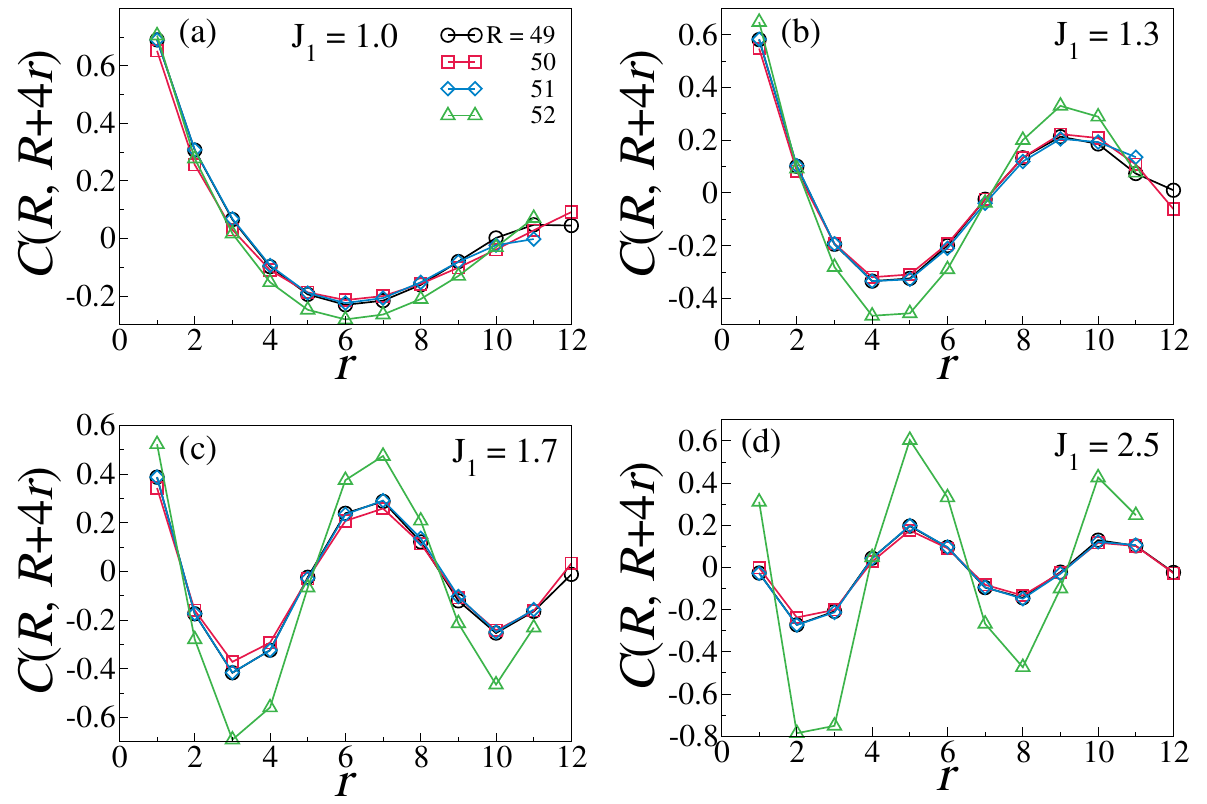}
        \caption{\label{fig7}Spin-spin correlations between spins at same type
	of sites in each unit cell of a 3/5 ladder of $N = 98$ spins
	(OBC) for (a) $J_1 = 1.0$, (b) $J_1 = 1.3$, (c) $J_1 = 1.7$, and (d)
	$J_1 = 2.5$. Starting from the middle of the ladder $R = 49$, 50, 51, 52
	are the reference sites.}
\end{figure}

To understand this seemingly strange behavior, we have computed spin densities
and spin-spin correlations for the gs with different spins (corresponding
to different $J_1$ values) for the $N = 98$ system. We see from Fig.~\ref{fig6}
that in the interior of the system, higher spin densities are found at $4k$
sites. Besides the spin densities of a given type of sites show a wavelike
behavior. Thus, the gs spin is dictated by two factors. Large $J_1$
favors a high spin gs, while the wavy nature of the spin density
favors a standing wave as interference effects will be lowest in this state and
so will be the spin fluctuations which tend to increase the energy of a state.
Hence, the gs shows the unusual spin
state changes as $J_1$ is increased. This also explains the nonmonotonic dependence on
the switching values of $J_1$ for different system sizes. When $J_1$ is very
large ($J_2 / J_1 \to 0$), we expect the gs spin of the system to
be $4n$ where $n$ is the number of unit cells in the system. This unusual behavior is
also reflected in the spin-spin correlation function (Fig.~\ref{fig7}), where
the wavelength of the correlation function decreases with $J_1$ and the
magnitude of the correlation between spins at sites $4k$ has larger amplitude
than the rest of the correlations. 

\begin{figure}
\includegraphics[width=\columnwidth]{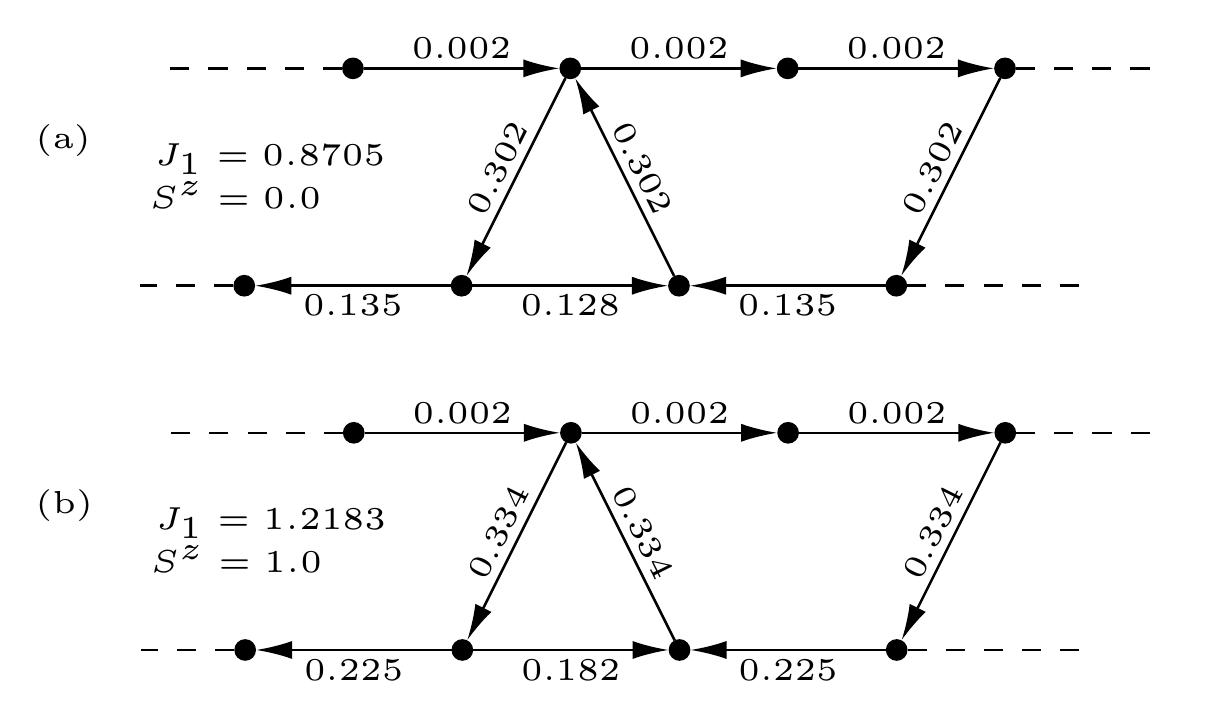}
\caption{\label{fig8}The spin current in a 3/5 skewed ladder of $N=16$ spins 
with periodic boundary condition at (a) $J_1 = 0.8705$ and (b) $J_1 = 1.2183$. The direction 
of the spin current is indicated by arrowheads and the magnitude is given by the numbers 
adjacent to the arrows.}
\end{figure}

In case of our 3/5 ladder system, there are two mirror planes one which is passing
through the centre of the five membered ring along the upper leg and
perpendicular to the lower leg, whereas second mirror plane `$\sigma$' passes
through the apex of triangle and bisects the base triangle. The system also has
spin inversion symmetry $P$ which exists in the $S^z = 0$ subspace and
corresponds to invariance of the Hamiltonian when all the spins in the system
are rotated by $\pi$ around the $y$-axis. This is equivalent to invariance of
Hamiltonian under spin inversion. This symmetry divides the $S^z = 0$ subspace
into an even (`$+$') and an odd (`$-$') subspace with the even (odd) subspace
spanning states with even (odd) spins. Thus, the even subspace consists of basis
functions with even integer total spin ($S$) and the odd subspace consists of
basis functions with odd integer total spin($S$). The symmetry group of the 3/5 
skewed ladder consists of  $E$, $P$, $\sigma$ and $\sigma P$ and all these 
elements commute with each other and form an abelian group. The irreducible 
representations of this group is represented by $A^+$ , $A^-$ , $B^+$ and 
$B^-$. $A(B)$ corresponds to even (odd) space under $\sigma$ and $+(-)$
corresponds to even(odd) space under $P$. The spin inversion symmetry is broken when the lowest energy
states with odd and even total spins are degenerate and reflection symmetry is
broken when lowest energy levels in both $A$ and $B$ spaces are degenerate.
Both the symmetries are broken when the lowest energy states in $A^+$ ($A^-$) and $B^-$ ($B^+$)
spaces are degenerate. We show the degeneracy of singlet and triplet states at
$J_1 = 0.8075$ and triplet and quintet at $J_1 = 1.2183$ in Table~\ref{tab1}. For these values of
$J_1$, the reflection symmetry is also broken as lowest energy states in spaces 
odd and even under reflections are also degenerate for these $J_1$ values. 
Hence we observe spontaneous spin current. The $z$-component of the spin 
current is defined as
\begin{eqnarray}
	\kappa^z(j,k) & = & -i \langle \Psi_G(-) \mid (\vec{S}_j \times \vec{S}_k)^z \mid
\Psi_G(+) \rangle \nonumber \\
	& = & \frac{1}{2} \langle \Psi_G(-) \mid (S_j^+ S_k^- - S_j^- S_k^+) \mid
\Psi_G(+) \rangle
\end{eqnarray}
which is the eigen value of the spin current operator $(\vec{S}_j \times \vec{S}_k)^z$
expressed as a matrix in the degenerate $\mid \Psi_G(+) \rangle$ and $\mid \Psi_G(-) \rangle$
basis
where the function $\mid \Psi_G(+) \rangle$ $\left(\mid \Psi_G(-) \rangle \right)$ is the
gs in the even (odd) subspace for reflection and even (odd) subspace for spin
inversion. In Fig.~\ref{fig8}, we show the spin currents for $J_1 = 0.8705$ and
$1.2183$. The current in triangle is counter clockwise, whereas it is clockwise 
in the pentagons and weak in upper leg of pentagons. Interestingly the 
spin current retains the qualitative features for both the $J_1$ values.
\begin{table}
\caption{\label{tab1}Lowest energy levels for different $S_z$ values of $N=16$}
\begin{ruledtabular}
\begin{tabular}{ c  c  c  c  c  c }
        $J_1$ &  $E(S^z = 0)$  &   $E(S^z = 1)$ &   $E(S^z = 2)$  \\ \hline
      0.8705  &  $-23.0172$    &   $-23.0172$                     \\
              &  $-23.0172$                                       \\  \hline
      1.2183  &  $-23.7554$    &   $-23.7554$   &   $-23.7554$    \\
              &  $-23.7554$    &   $-23.7554$                     \\ 
\end{tabular}
\end{ruledtabular}
\end{table}

\section{\label{sec4}Summary and conclusions}
In this paper we study the gs properties of spin-1 Heisenberg antiferromagnetic 
model on 3/4 and 3/5 skewed ladder geometries shown in Fig.~\ref{fig1}(b) and
(c). The 3/4 spin ladder system goes from a singlet state to partially
magnetized state. For $J_1 \geq 1.82$ the magnetization per unit cell is $\langle
m\rangle =2$ with each triangle in the system contributing $\langle m \rangle =
1$ to the magnetization in the gs. The gs of this system for $J_1 < 1.6$ is a
singlet and shows short range correlations. In this system there are two unique
sites and three unique bonds in the six site unit cell. In the weak $J_1$ limit,
the gs is a singlet and spin densities are zero and bond orders along the leg
are close to that of the  valance bond state of a spin-1 chain. The bond order
in the upper leg which connect the apex or base of two different triangles
decreases gradually as $J_1$ is increased while bond-order for base of the
triangle increases rapidly. In large $J_1$ limit or in magnetic gs the spin
densities for base and apex is $0.75$ and $-0.5$ respectively, and the effective
spin per triangle is 1. The bond order along the rung and base of the triangle
are $1.5$ or equivalent two spin-1/2 dimers and $-1$ for bond between a basal
site and an apical site, indicating a ferromagnetic interaction. In the large
$J_1$ limit, the effective spin one on each triangle interact ferromagnetically.
In the spin-1/2 skewed ladder system the ferromagnetic arrangement of spin is an
example of the McConnell mechanism~\cite{mcconnell63,geet} with
antiferromagnetic exchange (here $J_2$) between sites with positive and
negative spin densities leading to a ferromagnetic interaction between the
unpaired spins. The effective interaction between the spins in two neighboring
triangles can be written as $J_{\rm{eff}} = 2 J_2 \rho_a \rho_b$ where $\rho_a$
and $\rho_b$ are the spin densities at $a$ and $b$ type sites. We note that
ferromagnetic mechanism for spin-1 system seems similar to spin-1/2 system. 

The gs properties of the 3/5 ladder seems more complicated unlike the 3/4
system. The gs is a singlet for $J_1 < 0.84$ and shows spin density wave gs for
larger $J_1$ values. What is interesting is that the gs spin $S_G$ varies
between $0$, $1$ and $2$ before attaining the saturation value of $n$ (the
number of unit cells) for very large $J_1$. The switching of the gs depends on
system size and there is no apparent systematics. Analysis of spin densities and
spin-spin correlations seem to indicate that there is a competition between the
standing spin density wave of a lower spin state and the higher exchange
stability of the successively higher spin state. This seems to render a
qualitative explanation of the seemingly unsystematic switching in the gs
spin of the system as a function of system size. This system also shows
vector chiral phase due to simultaneous breaking of the reflection symmetry and
spin inversion symmetry. The broken symmetry phase is characterized by nonzero
spin currents in the system.

In summary, we have studied the gs properties of an antiferromagentic Heisenberg
spin-1 system on 3/4 and 3/5 skewed ladders. Both systems transition  from
singlet to a partially polarised gs on tuning $J_1$. The 3/5 system shows
switching of the spin state between singlet and different magnetic states
due to a competition between $J_1$ which favors a high spin state and standing
spin density wave which favors a lower spin state. For two different parameter
values the gs of the system has doubly degenerate gs which leads to breaking of spin parity
and reflection symmetry in the system resulting in spontaneous spin current. 
Although such systems have not yet been experimentally realized, we believe that
they can be realized in molecular magnets based on transition metal compounds.

\section*{Author's contributions}
SD and DD contributed equally to this work.

\begin{acknowledgments}
MK thanks SERB for financial support through project file No.
CRG/2020/000754. SR thanks Indian National Science Academy and DST-SERB for supporting this
work.
\end{acknowledgments}

\section*{Data Availability}
The data that support the findings of this study are available from the
corresponding author upon reasonable request.

%

\end{document}